\documentclass[12pt]{article}
\pdfoutput=1
\usepackage[utf8]{inputenc}
\usepackage{amsmath}
\usepackage{amsthm}
\usepackage{amsfonts}
\usepackage{amssymb}
\usepackage{mathrsfs}
\usepackage{textcomp}
\usepackage[french, english]{babel}
\usepackage{hyperref} 
\usepackage{courier}
\usepackage{graphicx}
\usepackage{caption}
\usepackage{subcaption}
\usepackage{xfrac}
\usepackage{array}
\usepackage{stmaryrd}
\usepackage{bbding}
\usepackage{empheq}
\usepackage{mathtools}
\usepackage{xfrac}
\usepackage{nicefrac}
\usepackage{geometry}
\usepackage{setspace}
\usepackage{siunitx}
\usepackage{parskip}
\usepackage[T1]{fontenc}
\usepackage{afterpage}

\usepackage[utf8]{inputenc} % allow utf-8 input
\usepackage[T1]{fontenc}    % use 8-bit T1 fonts
\usepackage{hyperref}       % hyperlinks
\usepackage{url}           
\usepackage{booktabs}       % professional-quality tables
\usepackage{amsfonts}       % blackboard math symbols
\usepackage{nicefrac}       % compact symbols for 1/2, etc.
\usepackage{microtype}
\usepackage{amsmath}
\usepackage{authblk} % simple URL typesetting
\usepackage{amssymb}
\usepackage{empheq}
\newcommand*\widefbox[1]{\fbox{\hspace{2em}#1\hspace{2em}}}

\usepackage{lipsum}
\usepackage{xargs}
\usepackage{xcolor}

\newcommand{\CC}{\mathcal{C}}

\DeclareMathOperator*{\argmin}{arg\,min}

\DeclareMathOperator{\Id}{Id}

\usepackage{lmodern}

\usepackage{empheq}
\usepackage{graphicx}

\usepackage{braket}
\usepackage{fdsymbol}

\definecolor{darkred}{rgb}{0.8, 0.0, 0.0}

\DeclarePairedDelimiter\norm{\lVert}{\rVert}%

\DeclareMathOperator{\Tr}{Tr}

\makeatletter
\def\blfootnote{\gdef\@thefnmark{}\@footnotetext}
\makeatother

\begin{document}

\title{Predicting Stock Returns with Batched AROW}

\author[1]{Rachid Guennouni Hassani$^{\clubsuit}$}
\author[2]{Alexis Gilles\textcolor{darkred}{$^{\vardiamondsuit}$}}
\author[2]{Emmanuel Lassalle\textcolor{darkred}{$^{\varheartsuit}$}}
\author[2]{Arthur D\'enouveaux$^{\spadesuit}$}

\affil[1]{\'Ecole Polytechnique}
\affil[2]{Machina Capital}

\blfootnote{$^{\clubsuit}$ rachid.guennouni-hassani@polytechnique.edu}
\blfootnote{\textcolor{darkred}{$^{\vardiamondsuit}$} agilles@machinacap.com}
\blfootnote{\textcolor{darkred}{$^{\varheartsuit}$} elassalle@machinacap.com}
\blfootnote{$^{\spadesuit}$ adenouveaux@machinacap.com}

\setlength{\marginparwidth}{2cm}

\maketitle

\begin{abstract}
\noindent We extend the AROW regression algorithm developed by Vaits and Crammer in \cite{vaits_re-adapting_2011}
to handle synchronous mini-batch updates and apply it to stock return prediction. By design, the model should be more robust
to noise and adapt better to non-stationarity compared to a simple rolling regression. 
We empirically show that the new model outperforms more classical approaches by backtesting
a strategy on S\&P500 stocks.
\end{abstract}

 \newpage

\section{Introduction}
Financial markets exhibit highly non-stationary behaviors, making it difficult to build predictive signals
that do not decay too rapidly (see \cite{Schmitt_2013,Cont01empiricalproperties} for empirical studies of
return time series). A standard
method for capturing these changes in time series data consists in using a rolling regression,
that is, a linear regression model trained on a rolling window and kept as static model
during a prediction period. However, the size of historical training data as well as the duration of the
prediction period have a direct impact on the performance of the resulting model: using too many training
data would result in a model that does not react quickly enough to sudden changes while short training
and prediction windows would make the model unstable (see for instance \cite{INOUE201755}).

Online learning algorithms are suited to situations where data arrives sequentially. New information
is taken into account by updating the model parameters in a supervised fashion. More precisely, an online learning algorithm repeats the following steps indefinitely: receive a new instance $x_t$,
make a prediction $\hat{y}_t$, receive the correct label $y_t$ for the instance and update the model
accordingly.

In the particular case of regression, online models are also good candidates to handle the non-stationarity inherent in financial time
series while keeping a certain memory of what has been learnt from the beginning. The recursive least squares (RLS) algorithm is a well known approach to online linear
regression problems (e.g. \cite{scharf1991statistical}), yet it updates the model parameters using one sample at a time. However, building predictive models on stock markets, one should take into account the very low signal over noise ratio,
and one way to do so is to fit a single model for predicting all the stock returns of a trading universe (hence fitting on much more data).
It allows us to have more training data covering shorter periods, but it comes with
the difficulty of updating the model parameters synchronously with all the data available at a given time.

In this paper we extend AROW for regression (\cite{vaits_re-adapting_2011}), an algorithm similar to RLS, in order to take into account a batch of instances for the online update instead of doing one update per sample. Indeed the latter approach would introduce a spurious order in information that actually occur synchronously and should be captured as such. Like RLS (see \cite{Hayes1996}), AROW suffers logarithmic regret in the stationary case, but also comes with a bound on regret in the general case. We test it on the universe of S\&P500 stocks on a daily strategy and show it outperforms the rolling regression on the same set of features in backtest.

\section{Adaptive Weight Regularization}

When training online models, we try to find the right balance between reactivity to new information and accumulation of predictive power over time. For instance, the family of models introduced in \cite{crammer_online_2006} aggressively update their parameters by guaranteeing a certain performance on new data (expressed in terms of geometric margin). But even regularized versions of these algorithms do not take into account the fact that some features might be less noisy than others. In other words, they are not suited for cases where we would like updates to be more aggressive on particular parameters than on the rest of the weight vector.

Subsequent online learning algorithms introduced in \cite{dredze_confidence-weighted_2008} and \cite{crammer_adaptive_2013} maintain a Gaussian distribution on weight vectors representing the confidence the model has in its parameters. From this point of view, the mean (resp. the variance) of the distribution
represents the knowledge (resp. the confidence in the parameters). In that framework, when receiving new feature vector and target $(x_t, y_t)$, we would like the Gaussian parameters to be updated by solving the following optimization problem:

\begin{align*}
(\mu_t,\Sigma_t) = \underset{\mu, \Sigma}\argmin \ &
D_{KL}\left(\mathcal{N}\left(\mu,\Sigma\right) \parallel
\,\mathcal{N}\left(\mu_{t-1},\Sigma_{t-1}\right)\right) \\
\quad s.t. \ & \mathbb{P}_{w \sim \mathcal{N}(\mu,\Sigma)}
\left(
\ell \left(y_{t},w^{\top} x_{t}\right)
\leqslant \epsilon
\right) \geqslant \eta
\end{align*}

where $D_{KL}$ is the Kullback-Leibler divergence and $\ell$ is the classical mean square loss. The idea behind this optimization is to find the minimal changes in knowledge and confidence such that minimum regression performance is achieved with a given probability threshold $\eta$.

Although that formulation is tractable (in particular there is an explicit formula for the KL divergence between two Gaussian distributions), it is not convex in $\mu$ and $\Sigma$. With AROW, Vaits and Crammer use a simpler but convex objective (\cite{vaits_re-adapting_2011}):

\begin{equation}\label{eq: mono update cost}
\mathcal{C}(\mu,\Sigma)=D_{KL}(\mathcal{N}(\mu,\Sigma) \parallel \mathcal{N}(\mu_{t-1},\Sigma_{t-1}))+\lambda_1 \ell(y_t,\mu^{\top} x_t) + \lambda_2 x_{t}^{\top} \Sigma x_t
\end{equation}
 
where $\lambda_1$ and $\lambda_2$ are hyperparameters to adjust the tradeoff between the three components of the objectives. These three components should be understood as follows:

\begin{enumerate}
    \item the parameters should not change much per update;
    \item the new model should perform well on the current instance;
    \item the uncertainty about the parameters should reduce as we get additional data.
\end{enumerate}

As such, the model is well suited for non-stationary regression. However the updates only take into account a single instance at the time. Because we want a single model for all the stocks here, we need to extend this approach to synchronously update the parameters on all cross-sectional information available at a given time, which is the purpose of the next section.

\section{Synchronous Batch Updates}

We now assume that between the updates at times $t$ and $t-1$, we have $K$ synchronous
instances
$(x_t^k, y_t^k)$, $k=1, \ldots, K$ of the features and targets. They correspond to the observation of a
complete universe of stocks at a given time. Applying an AROW update would introduce a fake order between the instances and potentially hurt the performance (see Section \ref{section:backtest} for more details).
In order to take into account all the new information at once in a single batch,
we extend the cost function from equation~\ref{eq: mono update cost} as follows:
\begin{equation*}
\begin{split}
\mathcal{C}(\mu,\Sigma)=&\,
D_{KL}(\mathcal{N}(\mu,\Sigma) \parallel \mathcal{N}(\mu_{t-1},\Sigma_{t-1}))\\
&+\frac{\lambda_1}{K}\sum_{k=1}^{K} \ell\left(y_t^k, \mu^{\top}x_t^{k}\right) + \frac{\lambda_2}{K} {x_t^{k}}^{\top}\Sigma x_t^{k}\,.
\end{split} 
\end{equation*}
For simplicity of presentation, we set $\lambda_1 = \lambda_2 = \frac{1}{2r}$ where $r > 0$ and $R = rK$. Using 
\begin{align*}
D_{KL}\Big(\mathcal{N}(\mu_1,\Sigma_1) \parallel \mathcal{N}(\mu_2,\Sigma_2)\Big) 
& = \frac{1}{2} \log \left(\frac{\det \Sigma_2}{\det \Sigma_1}\right) 
+ \Tr(\Sigma_2^{-1} \Sigma_1) \\
& \hspace{0.4cm} + \frac{1}{2} (\mu_2 - \mu_1)^{\top} \Sigma_2^{-1} 
(\mu_2 - \mu_1) - \frac{d}{2}\,,
\end{align*}
and setting 
\begin{equation*}
X_t=[x_t^{1} \cdots x_t^{K}]^{\top}\,, \quad
Y_t=[y_t^{1} \cdots y_t^{K}]^{\top} \,.
\end{equation*}
we get, remembering that $\ell$ is the mean square loss,
\begin{align*}
\mathcal{C}(\mu,\Sigma)=&\frac{1}{2}\log\left(\frac{\det \Sigma_{t-1}}{\det \Sigma}\right)
+\frac{1}{2} \Tr (\Sigma_{t-1}^{-1} \Sigma)+\frac{1}{2}(\mu_{t-1}-\mu)^{\top}\Sigma_{t-1}^{-1}(\mu_{t-1}-\mu) \\ & -\frac{d}{2}+\frac{1}{2R}\left(\norm{Y_t-\mu^{\top} X_t}^2 + \Tr(X_{t} \Sigma X_t^{\top})\right)\,.
\end{align*}

The main result of this article is the following:
minimizing the cost function $\CC$ has an explicit solution given by

\begin{subequations}
\begin{empheq}[box=\widefbox]{align}
\Sigma_t &= \Sigma_{t-1} - \Sigma_{t-1} X_t^{\top} 
\left(
R \Id_d + X_t \Sigma_{t-1} X_t^{\top}
\right)^{-1}
X_t \Sigma_{t-1}\label{them result1} \\
\mu_t &= \mu_{t-1} - \Sigma_{t-1} X_t^{\top}
\left(
R \Id_d + X_t \Sigma_{t-1} X_t^{\top}
\right)^{-1}
\left(
X_t \mu_{t-1} - Y_t
\right)\label{them result2}\,.
\end{empheq}
\end{subequations}

From now on we refer to this batch version of AROW as BAROW, and detail the steps to the solution in the next section.

\section{Deriving the Update Formulas}

As for AROW, BAROW's cost function $\CC$ is convex. To prove \ref{them result1} and \ref{them result2} it
is thus enough to look at the critical points of $\CC$.
We start by computing $\partial \CC / \partial \Sigma$ and, using formulas
from \cite{IMM2012-03274} chapter $2$, we find 
\[
    \frac{\partial \CC}{\partial \Sigma} (\mu, \Sigma) = 
    - \frac{1}{2} 
    \left(
    \Sigma^{-1} - \Sigma_{t-1}^{-1} - \frac{1}{R} X^{\top}_t X_t
    \right) \,.
\]
The Kailath variant of the Woodbury identity (see \cite{IMM2012-03274} chapter $3$) yields
\[
    \left(
        \Sigma_{t-1}^{-1} + \frac{1}{R} X_t^{\top} X_t
    \right)^{-1}
    =
    \Sigma_{t-1} - \Sigma_{t-1} X_t^{\top}
    \left(
        R \Id + X_t \Sigma_{t-1} X_t^{\top}
    \right)^{-1} X_t \Sigma_{t-1}
\]
and allows us to deduce that ${\partial \CC} / {\partial \Sigma}$ vanishes for
\[
    \Sigma = 
    \Sigma_{t-1} - \Sigma_{t-1} X_t^{\top}
    \left(
        R \Id + X_t \Sigma_{t-1} X_t^{\top} 
    \right)^{-1} X_t \Sigma_{t-1} \,,
\]
which is formula \ref{them result1}.

Similarly, one finds
\[ 
    \frac{\partial \CC}{\partial \mu} (\mu, \Sigma) =
    \Sigma_{t-1}^{-1}(\mu - \mu_{t-1}) 
    + \frac{1}{R} (X^{\top}_t X_t \mu - X^{\top}_t Y_t)\,,
\]
so that $\partial \CC / \partial \mu = 0$ if and only if
\begin{align*}
\label{eq: partial mu}
    \mu &= 
    \left(
        \Sigma^{-1}_{t-1} + \frac{1}{R} X^{\top}_t X_t
    \right)^{-1}
    \left(
        \Sigma^{-1}_{t-1} \mu_{t-1} + \frac{1}{R} X^{\top}_t Y_t
    \right) \\
    &=
    \left(
        \Sigma_{t-1}^{-1} + \frac{1}{R} X^{\top}_t X_t
    \right)^{-1} \Sigma_{t-1}^{-1} \mu_{t-1}
    +
    \frac{1}{R}
    \left(
        \Sigma_{t-1}^{-1} + \frac{1}{R} X^{\top}_t X_t
    \right)^{-1} X^{\top}_t Y_t
    \,.
\end{align*}
We deal with the two terms separately. For the first one
we use again the Kailath variant of the Woodbury equality and
get
\begin{align*}
    \left(
        \Sigma_{t-1}^{-1} + \frac{1}{R} X^{\top}_t X_t
    \right)^{-1} \Sigma_{t-1}^{-1} \mu_{t-1}
    &=
    \left(
    \Id - \Sigma_{t-1} X^{\top}_t
    \left(
        R \Id + X_t \Sigma_{t-1} X_t
    \right)^{-1} X_t
    \right) \mu_{t-1} \\
    &=
    \mu_{t-1} - 
    \Sigma_{t-1} X^{\top}_t
    \left(
        R \Id + X_t \Sigma_{t-1} X_t
    \right)^{-1} X_t \mu_{t-1} \,.
\end{align*}

For the second term, first notice that:
\begin{align*}
    \left(
        \Sigma^{-1}_{t-1} + \frac{1}{R} X^{\top}_t X_t
    \right)
    \Sigma_{t-1} X^{\top}_t
    &=
    \left(
        X_t^{\top} + \frac{1}{R} X^{\top}_t X_t \Sigma_{t-1} X_t^{\top}
    \right) \\
    &=
    \frac{1}{R} X_t^{\top} \left(R \Id + X_t \Sigma_{t-1} X_t^{\top}\right)
\end{align*}
so that
\[
    \left(
        \Sigma^{-1}_{t-1} + \frac{1}{R} X^{\top}_t X_t
    \right)
    \Sigma_{t-1} X^{\top}_t
    \left(
        R \Id + X_t \Sigma_{t-1} X_t^{\top}
    \right)^{-1} Y_t
\]
is equal to 
\[
    \frac{1}{R} X_t^{\top} \left(R \Id + X_t \Sigma_{t-1} X_t^{\top}\right)
    \left(
        R \Id + X_t \Sigma_{t-1} X_t^{\top}
    \right)^{-1} Y_t \,.
\]
Simplifying $(R\Id + X_t \Sigma_{t-1} X_t^{\top})$ we get
\[
    \frac{1}{R}
    \left(
        \Sigma^{-1}_{t-1} + \frac{1}{R} X^{\top}_t X_t
    \right)^{-1} X^{\top}_t Y_t
    = 
    \Sigma_{t-1} X_t^{\top}
    \left(
        R \Id + X_t \Sigma_{t-1} X_t^{\top}
    \right)^{-1} Y_t \,.
\]

Finally, combining the two terms, we get
\begin{align*}
    \mu 
    &=    
    \mu_{t-1} - 
    \Sigma_{t-1} X^{\top}_t
    \left(
        R \Id + X_t \Sigma_{t-1} X_t
    \right)^{-1} X_t \mu_{t-1}
    +
    \Sigma_{t-1} X_t^{\top}
    \left(
        R \Id + X_t \Sigma_{t-1} X_t^{\top}
    \right)^{-1} Y_t \\
    &=
    \mu_{t-1} -
    \Sigma_{t-1} X^{\top}_t
    \left(
        R \Id + X_t \Sigma_{t-1} X_t
    \right)^{-1} (X_t \mu_{t-1} - Y_t) \,,
\end{align*}
which was the claimed formula.

\section{Backtesting a Strategy using BAROW}\label{section:backtest}

BAROW combines the adaptability of AROW to non-stationary data and the advantage of taking into account all new information synchronously. We tested the model against two baselines as return predictor for a trading strategy on the S\&P500 universe.

We backtested a long-short strategy taking daily positions proportional to the prediction generated by a regression model on stock returns and showed it outperforms the following baselines:

\begin{enumerate}
    \item A rolling regression updated daily and using the past 12 months of data for training.
    \item AROW regression with single instance updates (500 updates per day).
\end{enumerate}

\begin{figure}[h] \label{fig:expected_pnl}
\begin{center}
\noindent
\makebox[\textwidth]{\includegraphics[scale=0.7]{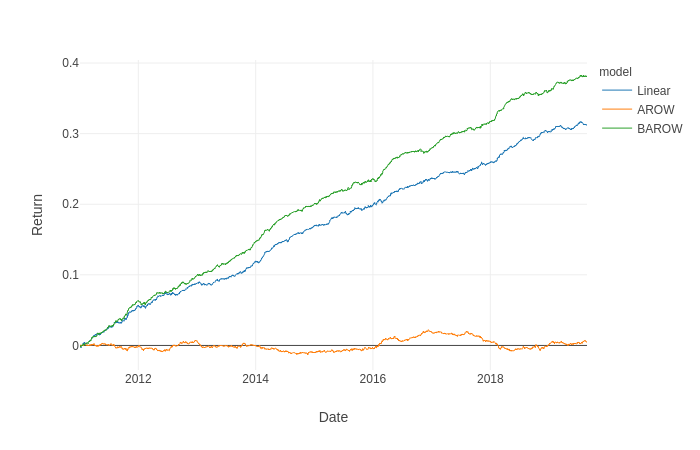}}
\caption{Estimated Return of each model}
\end{center}
\end{figure}

We ran the backtest on 2,250 days from 2011 to 2019, allowing a burn-in period of 12 months for AROW and BAROW before starting to use them as return predictors. We also tweaked the $R$ hyperparameter on 2010 for AROW and BAROW.
To avoid the strategy having too high sensitivity to market moves, we neutralized daily returns using a multi-factor model (including beta, volatility and a variety of other indicators, see \cite{Kakushadze_2015} for a detailed discussion about risk models). These neutralized returns are the regression targets $y_t$. We used features based on MACD indicators such as those described in \cite{Chong_2014}.

We estimated daily returns of a strategy taking positions proportional to the predictions as the cross-sectional correlation between the predictions and the realized returns, multiplied by the cross-sectional standard deviation of the returns.

As expected, we can see that the baseline using AROW updates stock per stock is not suited to learn a generic predictor for the universe. Its cumulative return appears random compared to the two other models. That empirically justifies the batch update computed in BAROW.

Now if we compare BAROW to the rolling linear regression, we directly see that performances drift away from each other. BAROW significantly outperforms the baselines, illustrating the benefits of online learning models for non-stationary data.

\begin{table}[h]
    \vspace{1cm}
    \centering
    \begin{tabular}{|c|c c c c|}
        \hline
        model & Return & Sharpe & MaxDD & Calmar \\
        \hline
         AROW & 0.5\% & 0.1 & -3.02\% & 0.18 \\
         Linear & 31.3\% & 4.0 & \textbf{-0.58\%} & 53.6\\
         BAROW & \textbf{38.1\%} & \textbf{5.1} & -0.61\% & \textbf{62.2} \\
        \hline 
        
    \end{tabular}
    \caption{Performance statistics of each model}
    \label{tab:stats}
\end{table}

We provide some usual performance statistics in table \ref{tab:stats}. Specifically, we report the total return over the period, the Sharpe ratio of expected return (\textit{Sharpe}), the maximum drawdown (\textit{MaxDD}), and the Calmar ratio (\textit{Calmar}, defined as return over the period divided by \textit{MaxDD}). One should be aware that using such short targets to predict induces a very high turnover in the portfolio, resulting in lower net performance after transaction costs are applied. 

\section{Discussion}

Our BAROW algorithm outperformed its baseline in a backtest, yet the question of how $\Sigma$ impacts the model updates remains. The model forces $\Sigma$ to converge, causing the model to be less and less able to adapt quickly. That trade-off between adaptability and robustness might not be the right one in volatility periods when one could require more dynamic updates. One way to address it is to schedule reset of the covariance matrix as experimented in \cite{vaits_re-adapting_2011}. In our case it might be beneficial to reset the covariance matrix conditionally on any market event we consider as a trigger for more dynamic updates.

The outperformance of BAROW vs its baseline and the adaptability/robustness trade-off mentioned above could both be further analyzed by increasing the backtest frequency (e.g. 1-min bars). A higher frequency would likely enable shorter training and prediction windows, hereby enhancing the potential overperformance of BAROW vs the linear baseline. 

Also, the backtest results described above are a mere sum of single-stock returns and do not include any form of risk management at the overall portfolio level, which could further enhance the risk-adjusted performance of each approach.

Finally, it is worth pointing out that BAROW results are significantly impacted by the method used to neutralize returns that serve as the regression target. Usual standardization methods have exhibited periods of higher correlation of residuals in recent years. These likely stem from stakeholders relying on similar risk models and an ever-spreading use of machine learning type of strategies.

\bibliographystyle{alpha}
\bibliography{references}

\newcommand{\etalchar}[1]{$^{#1}$}
\begin{thebibliography}{CDK{\etalchar{+}}06}

\bibitem[CDK{\etalchar{+}}06]{crammer_online_2006}
Koby Crammer, Ofer Dekel, Joseph Keshet, Shai Shalev{-}Shwartz, and Yoram
  Singer.
\newblock Online passive-aggressive algorithms.
\newblock {\em J. Mach. Learn. Res.}, 7:551--585, 2006.

\bibitem[CKD13]{crammer_adaptive_2013}
Koby Crammer, Alex Kulesza, and Mark Dredze.
\newblock Adaptive regularization of weight vectors.
\newblock {\em Machine Learning}, 91(2):155--187, May 2013.

\bibitem[CNL14]{Chong_2014}
Terence Chong, Wing-Kam Ng, and Venus Liew.
\newblock Revisiting the performance of macd and rsi oscillators.
\newblock {\em Journal of Risk and Financial Management}, 7(1):1–12, Feb
  2014.

\bibitem[Con01]{Cont01empiricalproperties}
Rama Cont.
\newblock Empirical properties of asset returns: stylized facts and statistical
  issues.
\newblock {\em Quantitative Finance}, 1:223--236, 2001.

\bibitem[DCP08]{dredze_confidence-weighted_2008}
Mark Dredze, Koby Crammer, and Fernando Pereira.
\newblock Confidence-weighted linear classification.
\newblock In {\em Proceedings of the 25th international conference on {Machine}
  learning - {ICML} '08}, pages 264--271, Helsinki, Finland, 2008. ACM Press.

\bibitem[Hay96]{Hayes1996}
Monson~H. Hayes.
\newblock {\em Statistical Digital Signal Processing and Modeling}.
\newblock John Wiley \& Sons, Inc., USA, 1st edition, 1996.

\bibitem[IJR17]{INOUE201755}
Atsushi Inoue, Lu~Jin, and Barbara Rossi.
\newblock Rolling window selection for out-of-sample forecasting with
  time-varying parameters.
\newblock {\em Journal of Econometrics}, 196(1):55 -- 67, 2017.

\bibitem[KL15]{Kakushadze_2015}
Zura Kakushadze and Jim Liew.
\newblock Custom v. standardized risk models.
\newblock {\em Risks}, 3(2):112–138, May 2015.

\bibitem[PP12]{IMM2012-03274}
K.~B. Petersen and M.~S. Pedersen.
\newblock The matrix cookbook, nov 2012.
\newblock Version 20121115.

\bibitem[SCSG13]{Schmitt_2013}
Thilo~A. Schmitt, Desislava Chetalova, Rudi Schäfer, and Thomas Guhr.
\newblock Non-stationarity in financial time series: Generic features and tail
  behavior.
\newblock {\em {EPL} (Europhysics Letters)}, 103(5):58003, sep 2013.

\bibitem[SD91]{scharf1991statistical}
L.L. Scharf and C.~Demeure.
\newblock {\em Statistical Signal Processing: Detection, Estimation, and Time
  Series Analysis}.
\newblock Addison-Wesley series in electrical and computer engineering.
  Addison-Wesley Publishing Company, 1991.

\bibitem[VC11]{vaits_re-adapting_2011}
Nina Vaits and Koby Crammer.
\newblock Re-adapting the {Regularization} of {Weights} for {Non}-stationary
  {Regression}.
\newblock In {\em Algorithmic Learning Theory}, pages 114--128, 2011.

\end{thebibliography}

\end{document}